\documentclass[10pt]{article}
\usepackage{aeguill}
\usepackage{graphicx}
\usepackage{epsf}

\usepackage{rotating}
\usepackage{overcite}
\usepackage{setspace}
\makeatletter 
\renewcommand{\@biblabel}[1]{\quad $^{#1}$} 
\makeatother 
\newcommand{\ina}{${I_{Na}}$} 
\newcommand{\ikr}{${I_{Kr}}$} 
\newcommand{\iks}{${I_{Ks}}$} 
\newcommand{\gks}{${G_{Ks}}$} 

\begin{document}



\title{
The role of M cells and the long QT syndrome in cardiac arrhythmias:
simulation studies of reentrant excitations using a detailed 
electrophysiological model}

\author{Herv\'e Henry and Wouter-Jan Rappel}

\date{Department of Physics \\ Center for Theoretical 
Biological Physics,\\ University of California, San Diego.\\
La Jolla, CA  92093-0319}

\maketitle
\begin{abstract}

In this numerical study, we investigate the role of intrinsic 
heterogeneities of cardiac tissue
due to M cells in the generation and maintenance of 
reentrant excitations using the 
detailed  Luo-Rudy dynamic model.
This model has been extended to include a description of the
long QT 3 syndrome, and is studied in both one  dimension,
corresponding to a cable traversing the ventricular wall,
and two dimensions, representing a transmural slice.
We focus on two possible mechanisms for the generation of reentrant
events.
We first investigate if early-after-depolarizations 
occurring in M cells can initiate reentry. 
We find that, even for large values of the long QT strength, 
the electrotonic coupling 
between neighboring cells prevents early-after-depolarizations 
from creating a reentry.
We then study whether M cell domains, with 
their slow repolarization, can function as wave blocks for premature
stimuli. 
We find that the inclusion of an M cell domain
can result in some cases in reentrant excitations and we determine 
the lifetime of the reentry 
as a function of the size and geometry of the domain 
and of the strength of the long QT syndrome.

\end{abstract}

\newpage

\textbf
{Spatial heterogeneity of cardiac tissue causes cells to
repolarize at different rates, leading to
a dispersion of repolarization time. 
Dispersion of repolarization  has been postulated 
to function as a substrate for reentry phenomena which occur 
when the propagation of the electric wave is blocked
in one direction causing the wave front to curl and reenter
the previously excited tissue.
The ensuing incoherent electrical activity of the heart
is thought to subsequently lead to ventricular fibrillation,
the leading  cause of sudden death in the
industrialized world.
In this numerical study, we investigate the role of intrinsic 
heterogeneities due to M cells in the generation 
and maintenance of reentrant excitations.
These cells, located in the midmyocardium, exhibit a 
prolonged action potential duration and the resulting 
dispersion of repolarization is exacerbated in patients with the 
long QT syndrome.
Using a detailed electrophysiological model,
two previously postulated mechanisms are investigated:
one in which the M cells create premature stimuli that 
lead to reentrant events and one in which the M cell domains 
function as wave blocks for electrically propagating waves
originating, as usually, from the endocardium.
We find that 
the first mechanism is unlikely to happen under the 
conditions we study, due to
the strong electrotonic coupling that is normally present
between neighboring cells.
The second mechanism, on the other hand, 
can result in reentrant excitations. We determine 
the lifetime of the reentry 
as a function of the size and geometry of the domain, 
and of the strength of the long QT syndrome.
}

\section{Introduction}

Spatial heterogeneity of cardiac cells is commonly believed to play 
a major role in the generation and maintenance of arrhythmia.
Both dynamic heterogeneity,  where the dynamical properties of the 
homogeneous tissue are responsible for the heterogeneity
\cite{Quetal00, Watetal01, Foxetala02}, and
intrinsic heterogeneity, due, for instance, to the presence of different 
types of cells in the cardiac tissue,  
can result in cardiac cells that
repolarize at different rates.
The resulting dispersion of repolarization can 
lead to regions that block propagating waves and can,
ultimately, result in reentry phenomena.

The focus of this study is intrinsic heterogeneities, which
are present in a variety of forms. 
One of  the most dramatic intrinsic heterogeneities is the
presence of a layer of cells, M cells, roughly located in the
middle of the myocardial wall \cite{SicAnt91, Antetal91}. 
This layer consists of cells that have distinct electrophysiological 
properties, most notably a prolonged action potential duration. 
The presence of these cells has been observed in many experimental studies  
and  has been found in the ventricles of the  
dog \cite{SicAnt91}, guinea pig \cite{Sicetal96} 
and human heart \cite{Droetal95}. 
M cells may constitute up to
40\% of the human left ventricular free wall \cite{Droetal95}
and the resulting action potential  prolongation
is most dramatic for isolated cells but remains
substantial in wedge preparations where the M-cells are
coupled through gap junctions to the cells in the epi- and
endocardium \cite{Yanetal98}.
The dispersion becomes even more dramatic under the influence of
cardiac drugs that prolong the action potential and/or in the
presence of slow pacing \cite{Antetal91}.
The ion channels responsible for the prolongation of
the action potential in M cells include  primarily a smaller slow activating
potassium current \iks, an augmented late
sodium current \ina\ and a larger sodium-calcium exchange 
current \cite{AntFis01}.

It has been hypothesized that the resulting transmural dispersion
plays a major role in the initiation and maintenance of
reentry \cite{AntSic94}.
In particular, the role of M cells in the initiation
of torsade de pointes, a polymorphic ventricular tachycardia,
in patients with the long QT syndrome has recently received
considerable attention \cite{Akaetal02}. 
The long QT syndrome is characterized by a prolonged ventricular
repolarization and is associated with a high risk of sudden cardiac death.
It manifests itself as an
abnormally long QT interval in the ECG and
can be either congenital, idiopathic or  acquired (for a
review see Ref. \citen{AntShi02}).
Several forms of long QT syndromes with  different 
cellular mechanisms and clinical characteristics have been identified.
Nearly all cardiac events in patients with long QT 1, associated with
a reduced \iks, are initiated during 
increased sympathetic activation induced by
physical or emotional stress \cite{Schetal01, Ant02}. 
On the other hand, in patients with  long QT 2,
characterized by a decrease in \ikr, 
and long QT 3, corresponding to a late \ina, 
cardiac events occur most often during rest or sleep \cite{Schetal01}.

Several possible mechanisms linking M cells and dispersion of repolarization
 with 
cardiac arrhythmias have been postulated.
In one, M cells develop early-after-depolarizations,
which are characterized by a depolarization at the end of the 
action potential plateau. This depolarization could then create a 
premature stimulus 
via the electrotonic interaction with neighboring cells. 
This premature stimulus will be partially 
blocked by the M cell domain but can initiate a wave 
in other directions. 
Once the repolarization is complete, this wave can 
reenter the M cell zone, leading to a classic figure-of-eight reentry 
\cite{Win91}.
Crucial to this mechanism is the ability for the early-after-depolarizations to excite neighboring cells
despite electrotonic effects that will tend to smooth out steep
gradients.

In another possible mechanism, the dispersion of repolarization due to M cells
creates regions with delayed repolarization that 
act as conduction blocks for subsequent waves. 
Under suitable conditions, the dispersion of repolarization 
can then lead to partial 
wave block and reentry. 
This scenario has recently been shown to be plausible 
in a canine wedge
preparation of the left ventricle \cite{Akaetal02}. 
In this study, a transmural cross section of the wedge
was visualized via voltage sensitive dyes and  
long QT 2 conditions were realized pharmacologically by
adding the \ikr\ blocker d-sotalol \cite{ShiAnt97}.
This led to a marked increase in dispersion of repolarization  
and the presence of distinct islands of M cells,  which resulted in areas 
with steep spatial
gradients of repolarization.
When the wedge was paced under bradycardic
conditions from the endocardium, followed by a
premature stimulus delivered at a short coupling interval 
(time interval between the premature stimulus and the previous,
regular stimulus),
the islands were found to function as
conduction blocks and torsade de pointes   ensued.
The underlying mechanism of the
resulting arrhythmia was demonstrated to be sustained intramural
reentrant excitations.

Only  a limited number of numerical studies investigating 
long QT syndromes, dispersion of repolarizations 
and its role in arrhythmias has been undertaken. 
Work by Rudy and colleagues investigated long QT syndrome 
in isolated cells
\cite{VisRud99,ClaRud01} and strands
of cells with embedded domains of M cells \cite{VisRud00,GimRud02}
but did not address the role of M cells in the initiation of reentry. 
Clayton \textit{et al.} \cite{Claetal01} studied the size of the vulnerable window, defined
as the time interval for which a premature stimulus 
initiates a unidirectional
wave, in homogeneous strands of tissue. 
They found that  
long QT conditions did not increase this size measurably.
However, they did not investigate propagation block in strands 
having domains consisting of electrophysiologically
different cells. 
They also investigated the tip trajectory in numerical models
for long-QT 1, 2 and 3, using a relatively 
large square homogeneous domain but did not study
the initiation of reentrant excitation.
 
In this paper, we use  a detailed electrophysiological
model \cite{LuoRud94} to investigate  the two possible mechanisms
detailed above. 
We first study if normally coupled tissue can generate 
unidirectional waves in one-dimensional (1D) strands  of tissue.
We then determine under what conditions a domain of M cells
will block propagation in a 1D cable.
We finally determine under what conditions a two-dimensional
(2D) sheet of 
tissue containing a layer of M cells will lead to reentrant 
excitation.
 
\vspace{1cm}
\noindent
\section{The Model}

To describe the electrophysiological properties of the cardiac cell we 
use the Luo-Rudy dynamic (LRd) model \cite{LuoRud94,FabRud00}. 
The dynamics of the tissue is described by the reaction-diffusion equation 
\begin{equation}
\frac{\partial V}{\partial t}= \frac{1}{ C_m \rho S_v} \nabla^2 V -
\frac{I_{ion}}{C_m}
\label{oned}
\end{equation}
where $V$ (mV) is the membrane potential,
$C_m$ ($\mu \rm{F}\, cm^{-2}$) is the membrane capacitance,
$\rho$ ($\Omega$cm) is the bulk resistivity, 
$S_v$ ($\rm{cm}^{-1}$) is the
surface to volume ratio and
$\nabla^2$ is the Laplacian operator. 
The combination of $\frac{1}{ C_m \rho S_v}$ gives the
diffusion constant $D$, which we have taken to be
$D=\frac{1}{ C_m \rho S_v}=0.11$ cm$^2$/s resulting in a planar wave 
speed of approximately $17$ cm/s. 
This speed is consistent with 
experimentally observed wavespeeds in canine ventricles
when measured perpendicular 
to fibers \cite{Preetal95}.
As is the case for all electrophysiological models,
the details of the LRd model are
contained in the description of the 
total ionic current $I_{ion}$ ($\mu \rm{A} \, cm^{-2}$)
flowing through the cardiac cell membrane.
In contrast to earlier models, which 
describe  only a small number of  ionic currents, the LRd model and other 
often so-called ``second generation'' models attempt to incorporate
as many significant currents as possible, including a more 
complete description of intracellular calcium dynamics.

Our choice of the LRd model was motivated by several considerations.
First, all other existing models we tested, including 
the first generation Luo-Rudy \cite{LuoRud91} and 
Beeler-Reuter \cite{BeeReu77} models and the
second generation model of Fox \textit{et al.} \cite{Foxetalb02}, 
exhibited spiral cores that were typically comparable to or larger
than the computational domain, taken to be typical of human ventricles.
Thus, these models are not suitable to investigate
sustained reentry in our computations.
Second, we are interested in modifying specific currents 
that are responsible for the distinct properties of M cells and 
that play a role in the long QT syndrome (see below).
The LRd model, being a second generation model,
allows us to change these currents directly. 

Since M cells are predominantly present in the midmyocardium,
we focus here on the effects of regional heterogeneities
in transmural geometries.
Furthermore, since the transmural wedge 
experiment in Ref. \citen{Akaetal02}
suggests that, at least in the initial stages, the reentry caused
by the long QT syndrome is a 2D phenomenon we have limited ourselves to
tissue sheets that represent thin transmural wedges.
We have taken the transmural dimension
to be 1 cm, while the other dimension is taken to be 
2.5 cm.
Also, we found it useful to investigate the behavior
of the tissue in a 1D cable that can be thought of as a straight line 
traversing the myocardial wall and perpendicular to the 
endo and epicardial walls. 
In all simulations, we have taken a spatial discretization of 
0.008 cm and a time step of 0.02 ms.
We have verified that using smaller time and space steps did 
not significantly modify the quantitative results.

\begin{figure}[t]
\begin{center}
\includegraphics{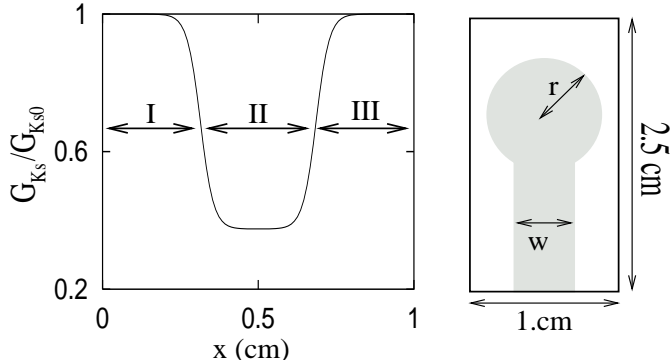}
\end{center}
\caption{
(a) Typical profile of the   
maximal conductance of \iks, $G_{Ks}$, 
used in the numerical studies of a cable. 
The conductance is shown normalized by the 
maximal conductance in normal tissue, $G_{Ks0}$. 
The cable consists of a region of width $w$ with a reduced
$G_{Ks}$ (II) corresponding to M cells, sandwiched between two 
equally sized domains of normal tissue (I and III).
(b) The geometry considered in our 2D simulations, consisting of 
a region of M cells, shaded in gray, surrounded by normal tissue.
As in the cable, the M cells are modeled via the reduction of 
$G_{Ks}$ and the transition between the different domains is
taken to be smooth.
}
\label{geom}
\end{figure}

Spatial heterogeneity was introduced by subdividing the cable into 
three domains (see Fig. \ref{geom}a). 
Since we focus here primarily on the role of the M cell domain,
we have chosen two outer domains that have equal length and
identical electrophysiological properties. 
This is, of course, a simplification 
as the endo- and epicardium  can have different
electrophysiological properties\cite{AntFis01}.
However, in the canine wedge preparation, the action potential 
durations of the epi- and 
endocardial cells were only  minimally  different \cite{Akaetal02}.
The middle domain consists of a region of M cells, which, 
electrophysiologically, were modeled  
by decreasing the value of the
maximum conductance \gks\ of the slow activating 
potassium current \iks. 
We tuned the value of \gks\ such that 
the epi- and endocardial cells and the M cells had 
an action potential that correspond to the experimental findings in 
Ref. \citen{Akaetal02}. Specifically, we took the normal tissue to have 
a maximal conductance \gks$_0$
that was a factor of 1.5 larger than the original value in the 
LRd model while the  maximal conductance \gks$_1$ in the M cells was
chosen as $G_{Ks1}=0.375 G_{Ks0}$. 

The transition between the two regions was made smooth 
by employing a simple sigmoidal function for \gks\ 
(see Fig. \ref{geom}a and supplemental material).
The two-dimensional (2D) geometry consists of a transmural
slice of the myocardial wall containing a strip of M cells.
Of course, it is impossible to initiate reentry via uniform
pacing of the entire endocardium
if the computational domain is  
invariant under translation along the endocardium 
(the vertical direction in Fig. \ref{oned}b).
Thus, to break the symmetry we have chosen 
to use an M cell layer consisting of a strip of width $w$ that
starts at one boundary and ends with a ``bubble" with radius $r$
at the other end
(see Fig. \ref{geom}b).
Evidence for inhomogeneous distributions of M cells
can be  seen in ref.
\cite{Akaetal02}  where the repolarization maps show distinct
islands of slower repolarization, presumably corresponding to
M cell domains.
As in the 1D cable, the transition between the different domains was made 
smooth.

We will focus here mostly on long QT 3, which is characterized by an
incomplete inactivation of the sodium current \ina\ and which leads 
to an abnormally prolonged action potential, particularly in M cells. 
To model the \ina\  channel and to allow  a description of the 
long QT 3  mutant we have adopted the strategy of Clancy and Rudy 
and have replaced \ina\ in the LRd model with a Markovian description 
\cite{ClaRud99}.
That is, instead of calculating the opening and closing of the 
fast sodium channel using  gating variables that
are deterministic functions of the membrane potential, we 
describe  this channel  using  a collection of open, closed and 
inactivation states. The transition rates between these states 
are dynamically calculated and are functions of $V$.
In addition to the background mode, also present in wild type cells,
mutant cells have a burst mode which does not include inactivation states.
When the channel is in this burst mode, the inactivation transiently fails,
leading to a leak current and a prolonged action potential.

To simplify the original model for \ina\ \cite{ClaRud99} and its subsequent
refinement \cite{ClaRud02}, and to render our computations 
more efficient, we have changed the original Markovian model. 
More details of this modification can be found in the supplemental material.
In essence, the simplification uses the existence of two distinct timescales 
of the \ina\ channel: a short timescale on the order of a single 
action potential and 
a much longer timescale on the order of multiple action potentials. 
Using these two timescales the Markov model can be reduced to 
an equation for \ina\ which is similar to the original formulation 
in the LRd model supplemented by a dynamical equation which describes the
fraction of channels in the burst mode.
Numerical checks, presented in the supplemental material, show
that the reduced description 
accurately reproduces the leak current.
Furthermore, we have verified that action potentials 
generated with the simplified model do not differ significantly from those 
 generated using the full model.

The simplification of \ina, along with the geometry of our computational
domain, lead to several adjustable parameters in our simulations. First, 
the strength of the long QT syndrome can be changed via a single 
dimensionless parameter $\mu$. 
It is a measure of the fraction of mutant cells (see appendix) and 
is equal to zero in the case of normal cells.
Second, the width of the M cell layer $w$ can be changed.
Finally, in 2D we have the additional parameter $r$, the 
radius of the disk at the end of the M cell layer. 
Notice that the minimal value of $r$, $r=w/2$, corresponds to a strip
with a rounded edge.
  
\section{Results}

\begin{figure}
\begin{center}
\includegraphics{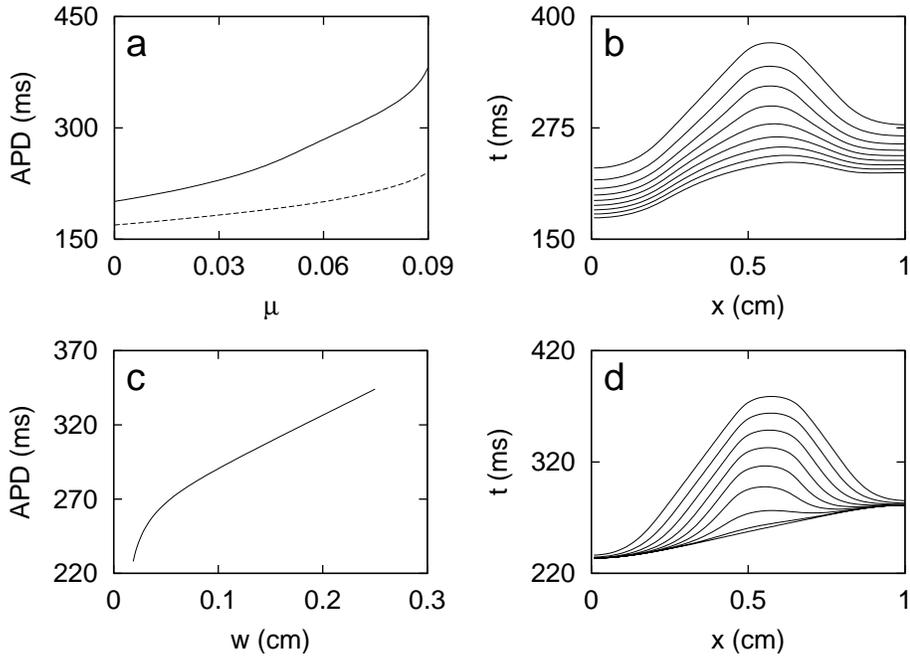}
\caption{
\label{apds1d}
(a): Action potential duration at a basic cycle length of 1000 ms 
as a function of the strength of the leak 
\ina, $\mu$, at the midpoint of a cable
with (solid line, $w=0.48$ cm) and without an M cell 
domain (dashed line).
(b): Corresponding repolarization times
for different values of $\mu$ varying from 0 (bottom
line) to 0.09 with steps of 0.01. 
(c): Action potential duration  at the midpoint of the cable 
as a function of 
the width of the M cell domain  (basic cycle length=1000 ms and $\mu=0.08$).
(d): Corresponding repolarization time 
for different widths  of the M cells region 
ranging from a homogeneous cable ($w=0$ cm and bottom line)   
to a domain of width $w=0.48$ cm with increments of 0.06 cm.}
\end{center}
\end{figure}

\subsection*{One dimension} 

We first quantified the effect of the size of the M cell domain $w$ and 
of the long QT syndrome strength $\mu$ on action potential duration 
in a cable.  For this, we stimulated the cable at one end with a fixed 
basic cycle length of 1000 ms for 40 s.
We then measured the action potential duration in the cell at the 
midpoint of the cable (and thus at the midpoint of the M cell domain)
and at the end of the cable.
In Fig. \ref{apds1d}a we plot
the action potential duration 
as a function of $\mu$ at the midpoint of 
a cable with (solid line, $w=0.48$ cm) and 
without an M cell domain (dashed line, $w=0.0$ cm).  
Fig. \ref{apds1d}b shows the corresponding repolarization times
along the cable, measured as the time interval between the 
stimulus and the time at which the membrane potential has 
repolarized to $V=-70$ mV.
As expected, increasing the long QT syndrome 
strength results in an increase in action potential durations. 
Since the action potential duration in M cells is more strongly 
affected by an increase in $\mu$,  higher values of $\mu$ result 
in an increased dispersion of repolarization time in the tissue. 
In Fig. \ref{apds1d}c we show  the action potential durations 
at the midpoint of the heterogeneous cable as a function 
of $w$ for  $\mu=0.08$, the long QT syndrome 
strength employed in most of this study.
The corresponding repolarization times can be found in 
Fig. \ref{apds1d}d.
The action potential duration of points well into the  normal tissue are hardly affected
by increasing $\mu$ while the midpoint of the cable displays a 
marked increase in action potential duration 
when $w$ is increased. The action potential durations computed numerically 
here   for $\mu=0$ and  $\mu=0.08$ are similar to the ones observed
in Ref. \citen{Akaetal02} when the wedge was paced 
under bradycardic conditions with and without d-sotalol, respectively.

\begin{figure}
\begin{center}
\includegraphics{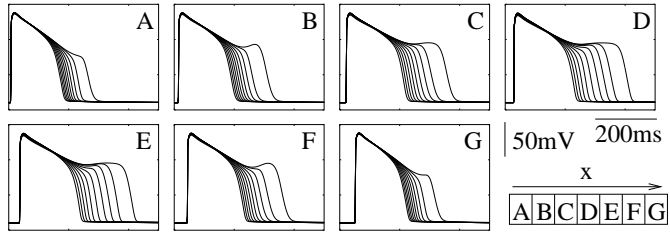}
\caption{\label{apcable} Action potentials at different positions along a cable 
of length 1 cm for increasing values of $\mu$ from 0 (shortest action potential)
to 0.09 (longest action potential) with steps of 0.01.  
The domain of M cells has a width of $w=0.48$ cm and corresponds to the region 
c, d and e, while the transition regions are b and f. 
The basic cycle length is 1000 ms, the 
horizontal bar corresponds to 
200 ms and the vertical bar corresponds to 50 mV.
}
\end{center} 
\end{figure} 

The shape of the action potential at different sites of the cable can be 
seen in Fig. \ref{apcable}. Here, we have divided the cable into 
7 subdomains and have plotted the action potential for each subdomain
for a fixed stimulation period of 
1000 ms and varying long QT syndrome strength.
The domain of M cells has a width of $w=0.48$ cm 
and corresponds to the region c, d and e.
It can clearly be seen that the long QT syndrome 
has a more dramatic effect on the M cells
(d) than on normal cells (a) and (g). 
Also note that due to the symmetry breaking introduced by the 
pacing at the left end of the cable, the action potentials 
are longer at the right boundary of 
the M cell domain (f) than at 
the left boundary (b). 

\begin{figure}
\includegraphics{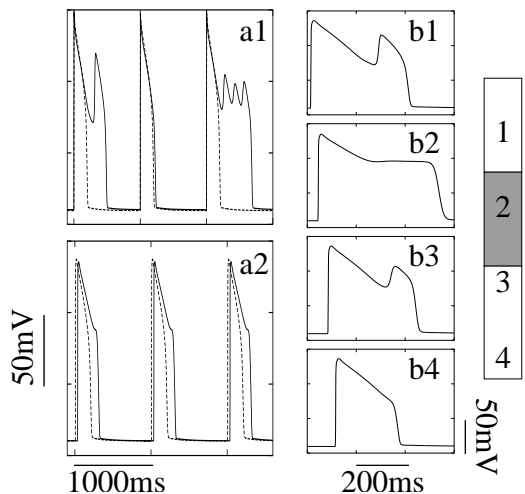}
\caption{\label{figEADs}
a1: The membrane  potential for a single isolated M cell (solid line)
for a long QT strength $\mu=0.06$,
along with the membrane  potential for a single isolated
epi- or endocardial cell (dashed line).
The cell is paced with a basic cycle length equal to 1000 ms.
a2: Membrane potential at different positions along our 
inhomogeneous  cable  ($w$=0.48 cm)  
consisting of cells used in a1.
The solid line correspond to the midpoint of the M cell domain, while
the dashed line correspond to the middle of the epicardial domain. 
b1-b4: The membrane potential at four different locations along 
the cable, as indicated by the graph on the right. 
The M cell domain ($w$=0.48 cm) is shaded gray and the cable
($\mu$=0.09) is paced using a basic cycle length of 
1000 ms, followed by a stimulus with a coupling interval of  
2000 ms.
}
\end{figure}

Next, we investigated if the presence of M cells could provoke 
early after depolarizations 
and waves propagating unidirectionally in our cable. 
We first verified that single M cells exhibit  
early after depolarizations, defined as an increase
in the membrane potential during the plateau phase of the action potential.
We found that these early after depolarizations are  present 
for a relatively wide range of parameters and pacing protocols
and have an amplitude that is consistent with experimental findings.
This is shown in Fig. \ref{figEADs}a1 where
we plot, as a solid line,
three subsequent action potentials for an isolated M cell with $\mu=$0.06. 
In contrast, an isolated epi- or endocardial cell, paced at the 
same basic cycle length of 1000 ms, does not exhibit 
early after depolarizations (dashed line).

Having established the presence of early after depolarizations in isolated
cells, we next considered a cable containing a domain of M cells.
In Fig. \ref{figEADs}a2, we show the action potential
in a cable consisting of cells identical to the ones of 
Fig. \ref{figEADs}a1. The solid line represents  
the membrane potential at
the middle of the M cell domain ($w=0.48$ cm), while the dashed
line represents the membrane potential 
in the middle of 
the epicardial domain. 
The early after depolarizations, present in the isolated cells, 
have now disappeared due to electrotonic effects. 
We found qualitatively similar results for different basic cycle lengths, 
long QT strengths and widths of the M cell domain:
early after depolarizations provoked in single M cells were always
found to be suppressed by electrotonic effects once the 
cells were inserted in a cable containing an M cell domain sandwiched
between an endo- and epicardial domain.

In fact, we found that, for the model used here, the only 
possible way to generate an early after depolarization that
travels along a homogeneously coupled
cable is to either change the 
pacing protocol or to increase the long QT strength significantly. 
Both scenarios lead to early after depolarizations that 
are already present in the endocardial domain.   
An example is shown in 
In Fig. \ref{figEADs}b where we have plotted the 
action potential at four different locations along the cable
as indicated in the figure.
The cable was paced at a constant basic cycle length 
of 1000 ms, followed by a stimulus with coupling interval 
of 2000 ms.
The endocardial domain now generates an 
early after depolarization with a 
significant amplitude that is able to travel 
through the M cell domain into the epicardial domain. 
However, the early after depolarization is propagating with a
smaller speed that the repolarization wave and 
disappears before reaching the end of the cable. 
Thus, in conclusion, 
under normal coupling conditions, the diffusive term in Eq. 
(\ref{oned}) always leads to considerable dissipation of the membrane
potential, which prevents the early after depolarizations 
from stimulating neighboring tissue.

\begin{figure}
\begin{center}
\includegraphics{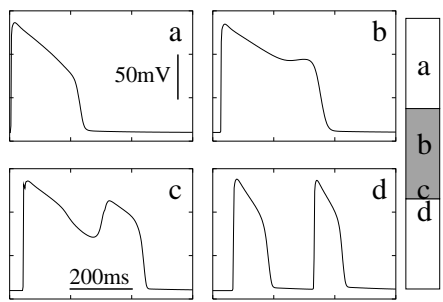}
\caption{\label{discon}
Action potentials at different positions along a cable
of length 1 cm shown on the right. 
The shaded M cell domain  has width $w=0.6$ cm and the 
cable ($\mu=0.08$) is paced with basic cycle length of 1000 ms, followed
by a stimulus with coupling interval of 1200 ms. 
The maximal conductance  \gks\ of the 
slow activating potassium current was increased by a factor of 
two in the epicardial domain.
The resistance between the last gridpoint of the
M cell domain and the first gridpoint of
the epicardium was abruptly increased by a factor of thirteen and
returned to normal over the following ten gridpoints.
}
\end{center}
\end{figure}

Some studies have suggested that a discontinuity in conductivity
exists between the M cell domain and the epicardium, leading to
inhomogeneous coupling \cite{Yanetal98}. 
It is possible that the inclusion of such inhomogeneous coupling, as
shown in other modeling studies that investigated
the propagation from the Purkinje fibers
(specialized fibers 
that rapidly transmit impulses from the 
atrioventricular node to the ventricles)
to the mycardium \cite{Saietal99, Monetal00},
allows for unidirectional conduction.
However, we found that 
propagation provoked by early after depolarizations in
our model was only possible when we altered
the conductance significantly. 
Specifically, we found that we had to reduce the  
conductance by at least a factor of ten, which 
is much more that the factor of three reported in Ref. \citen{Yanetal98}.
An example is shown in  Fig. \ref{discon} where we have plotted 
the membrane potential at four different locations in the cable,
as indicated in the figure. 
An early after depolarization appears in the 
M cell domain (b) as the coupling interval is increased.
This early after depolarization becomes more pronounced at the boundary
of the M cell domain and the epicardium, 
where the resistance is increased by a factor of thirteen. 
The increased resistance allows the epicardium to repolarize and the 
early after depolarization eventually creates 
an extra stimulus in the epicardium.

These results suggest that the first arrhythmic mechanism described in  
the Introduction is unlikely to occur in the model 
we consider here. 
To study the second mechanism, we recall that 
this scenario predicts the blocking of a wavefront by the 
remnants of previous stimuli. 
To investigate the possibility of a wave block created by the M cell 
domain, we again paced the cable at one end with a fixed period.
After 40 cycles, we 
delivered a premature stimulus to the same end of the cable
and varied the coupling interval
between the last of the regular cycles 
and the premature stimulus.
Conduction block was defined as an 
inability to propagate through the M cell domain to the opposite end
of the cable. 
The experiment was repeated for different  basic cycle lengths
and the conduction block window, the range of coupling intervals that
resulted in a conduction block, was determined.

\begin{figure}
\begin{center}
\includegraphics{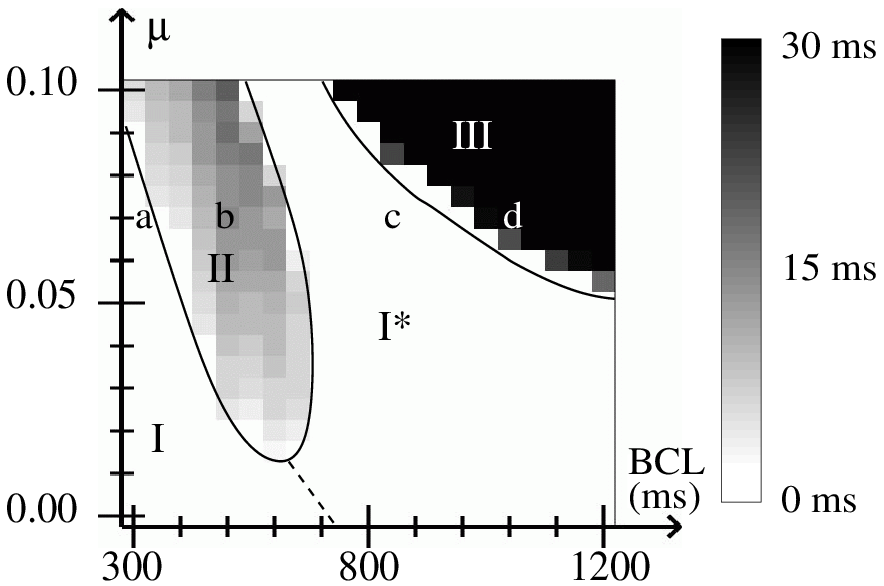}
\caption{
Phase diagram of the response of a cable paced at a 
regular basic cycle length (BCL)
to a premature stimulus in the presence of a M cell domain of width 0.48 cm
and of varying long QT syndrome strength.
In regions I and I$^{*}$ no conduction block occurs: 
a wave that has begun to propagate 
will propagate through the cable.
In region I the wave speed in the M cell domain is 
similar to the wave speed in the normal domain while in 
region $\mbox{I}^*$ the wave in the M cell domain 
slows down considerably before resuming normal propagation.
The dashed line is a guide to the eye to separate the two different
regions.
Regions II and III correspond to two different types of conduction block. 
In region II the normally propagating wave
reaches the M cell region and fails to propagate almost immediately.
Region III, on the other hand, corresponds to conduction block in which
the wave is able to enter the M cell domain, where it propagates with 
a significantly decreased wave speed before failing to propagate.
The size of the window of coupling intervals which lead to conduction 
block in region II and III is indicated by a gray scale, with
white being the smallest range and black being the largest range. 
\label{phase1d} }
\end{center}
\end{figure}

\begin{figure}
\begin{center}
\includegraphics{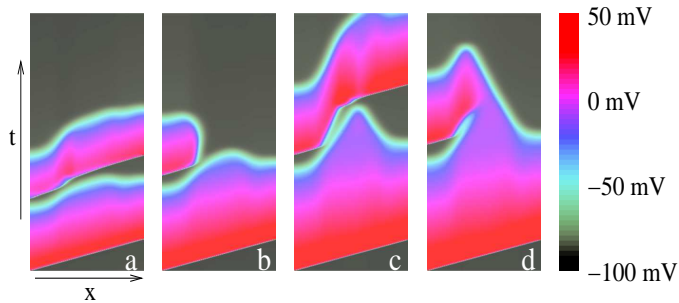}
\caption{
Space-time diagrams of the transmembrane potential of a cable 
of length 1 cm that includes an M cell domain of 
width $w=0.48$ cm and long QT strength $\mu=0.07$.
The cable is paced at the left end 
with constant basic cycle length for 10 s, after which a 
premature stimulus of varying coupling intervals is given. 
The color code used is shown by the colorbar on the right.
The four figures a-d correspond to 
the points marked a-d in Fig. \ref{phase1d} and are typical of the 
dynamics in the four different regions in that figure.
(a) Dynamics representative of region I in Fig. \ref{phase1d}: the 
premature stimulus (coupling interval=140 ms)
propagates through the M cell domain with almost the 
same speed (basic cycle length=300 ms).
(b) Dynamics Representative of region II in Fig. \ref{phase1d}: the 
wave is blocked immediately when entered the M cell domain 
(basic cycle length=500 ms
and coupling interval=184 ms).
For this basic cycle length, the conduction block window
 was fairly small (15 ms) with
conduction block occurring for a coupling interval between 170 ms and 
185 ms. 
(c) Dynamics Representative of region $\mbox {I}^{*}$ in Fig. \ref{phase1d}:
due to the slow repolarization of the M cells the wave speed is 
significantly decreased when entering the M cell domain. 
The speed increases again when
the wave can propagate 
freely through the repolarized medium (basic cycle length=800 ms 
and coupling interval=224 ms).
(d) Dynamics Representative of region III in Fig. \ref{phase1d}:
the wave speed decreases dramatically upon entering the 
M cell domain. However, unlike (c), normal propagation is unable to 
resume and the wave is blocked (basic cycle length=1000 ms and 
coupling interval= 245 ms).
For this stimulation period, we found a conduction block window 
of 72 ms with conduction block occurring for a coupling interval
 between 225 ms and 297 ms.
\label{spacetime1d}}
\end{center}
\end{figure}

 We found a rich set of possible dynamics that is 
summarized in Fig. \ref{phase1d}
where we have plotted our findings in the basic cycle length
(BCL) vs. strength of the long QT syndrome ($\mu$) phase
space. 
For wild type cells, i.e. $\mu=0$, the presence of the M cells domain does not 
lead to any conduction block. 
However, qualitatively different wave 
propagations are present.
For small pacing periods, the premature wave travels through 
the M cell region with almost unchanged speed. 
This behavior is demonstrated 
in Fig. \ref{spacetime1d}a which shows a color-coded space-time plot 
of the membrane voltage. 
The parameter values for Fig. \ref{spacetime1d}a correspond to 
the point marked a in Fig. \ref{phase1d}.
Its behavior is typical of all points in 
region I in Fig. \ref{phase1d}.
For longer pacing periods, the wave front slows down significantly
when entering the M cell region. 
After coming almost completely to a stop, with typical 
wave speeds found to be around 4 cm/s, the wave 
resumes its original speed and propagates into the 
epicardium. 
This slow-down-and-go behavior is demonstrated
in Fig. \ref{spacetime1d}c, corresponding to point c in 
Fig. \ref{phase1d}, and is representative of the points in 
the region labeled I$^*$ in Fig. \ref{phase1d}.

For sufficiently large values of $\mu$, conduction block occurs
(region II and III in Fig. \ref{phase1d}).
Two qualitatively different types of conduction block are observed. 
In region II, the wave penetrates the M cell domain only slightly before 
being blocked. 
A space-time
plot of point b within region II is shown in Fig. \ref{spacetime1d}b.
The conduction block window for points in region II is
rather small, typically less than 20 ms.
This range is indicated in  Fig. \ref{phase1d} on a linear gray scale
with white corresponding to small conduction block windows and black
corresponding to large ones ($>$ 30 ms).   
The conduction block window for point b was found to be 15 ms.

\begin{figure}
\includegraphics{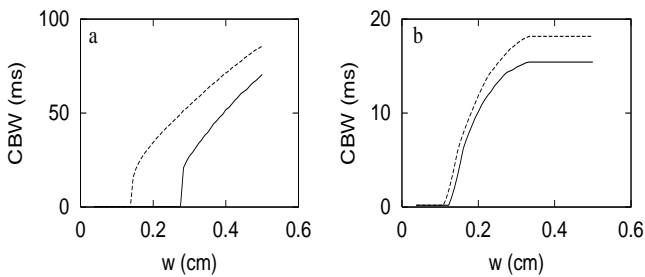}
\caption{The conduction block window duration (CBW) as a function of the width of the M cell domain  for $\mu=0.07$ (dashed lines) 
and $\mu=0.08$ (solid lines) for  a basic cycle length of 1000 ms (a) and 500 ms (b). Note the different scales of the CBW axis.
}
\label{cbw}
\end{figure}

The second type of conduction block can be found in region III and is shown 
in Fig. \ref{spacetime1d}d.
A wave front enters the M cell domain, 
reduces its speed significantly and is subsequently blocked.
The conduction block window for this region is much 
larger than for region II.
For example, the conduction block window  for point d is 72 ms.
This is also shown in Fig. \ref{cbw} where we have plotted the 
conduction block 
window as a function of the width of the M cell domain for two 
different values of the long QT strength and for two different 
values of the pacing period.
For both regions II and III  the conduction block windows 
increases as $w$ is increased.
We have also investigated the location of the boundaries of the regions in 
Fig. \ref{phase1d} for different $w$.  
Increasing $w$ did not significantly increase the size of any of the 
regions, while decreasing $w$ reduced the size of region II and III. 
Of course, for $w=0$ no conduction block occurs and
both region II and III are absent in the phase diagram.

In most of region III the behavior of the wave
as a function of the coupling interval for fixed basic cycle length 
and $\mu$ can 
be summarized as follows: 
For large coupling intervals, the wave is able to propagate 
normally through the M cell domain. 
Upon decreasing the coupling interval 
the wave slows down significantly
inside the M cell domain. Below a critical coupling interval 
conduction stops within the M cell domain.
Finally, after decreasing the coupling interval even further,
propagation fails already in the epicardial domain (domain I in 
Fig. \ref{geom}a).
Close to the boundary of region III with region I* 
a different behavior is observed. 
Again, for large coupling intervals the wave is able to propagate
normally through the M cell domain. For smaller coupling intervals
the wave slows down, and below a first critical coupling interval, 
propagation fails. 
Upon further decreasing the coupling interval, however, 
wave propagation resumes, characterized by a slowly propagating 
front within the M cell domain. 
Below a second critical coupling interval 
propagation fails in the epicardial domain. 
We will leave the discussion of this behavior to future studies.

In summary, our 1D results show that, for properly timed premature stimuli and 
sufficiently large values of $\mu$, the presence of a domain 
of M cells can create a conduction block.
Furthermore, early-after-depolarizations were observed 
to develop within the M cell domain.
However, in our numerical experiments, 
they are unable to excite already repolarized 
neighboring tissue due to electrotonic interactions.
Thus, reentrant excitations will not 
be generated in our model via the first mechanism 
and in our 2D simulations we will exclusively
focus on the second mechanism. 

\subsection*{Two dimensions} 

To investigate whether the second mechanism could lead to 
reentrant excitation in 2D we performed an extensive numerical 
study of a transmural sheet, shown in Fig. \ref{geom}b.
Our simulation protocol was chosen to reduce the computational cost 
while minimizing any drift in the action potential duration.
For this, we first divide the 2D sheet into 1D cables running from the endo-
to epicardium. These cables are then paced from the 
endocardial end for 40 s, after which 
the sheet is reassembled. Then,  the entire sheet is paced 
for a further 10 s by stimulating the entire endocardium.
This is finally followed by a premature stimulus
with a varying coupling interval.
Rather than choosing the premature stimulus at the epicardium
as was done in Ref. \citen{Akaetal02}, we
used a stimulus originating from the entire endocardium.
This can be thought of as a premature stimulus
coming from the Purkinje fiber \cite{Monetal00}. 

\begin{figure}
\begin{center}
\includegraphics{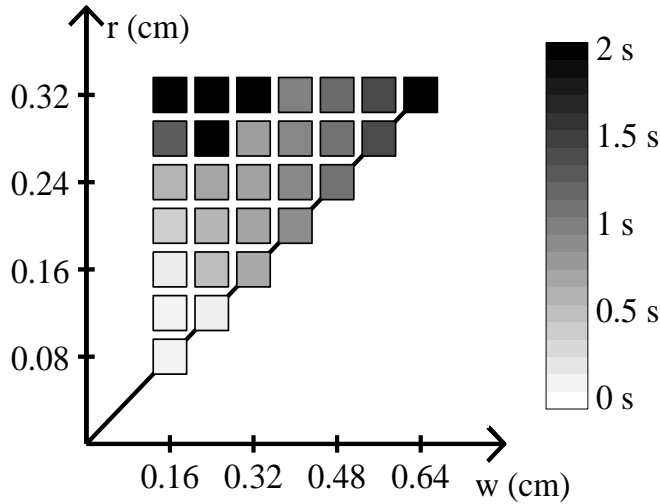}
\caption{
Phase diagram: Duration  of
the reentry as a function of the two parameters governing the geometry of 
the 2D sheet: the width of the M cell region ($w$) and the radius of the 
bubble ($r)$.
The gray-scale indicates the duration as shown by the bar on the right 
(units are in seconds). 
A relatively sharp transition occurs
between a regime for which 
reentries last at most around 1 second and a regime for which reentries last 
for more than two seconds and are pinned by the bubble (black squares).
\label{2dphase}}
\end{center}
\end{figure}

\begin{figure}
\begin{center}
\includegraphics{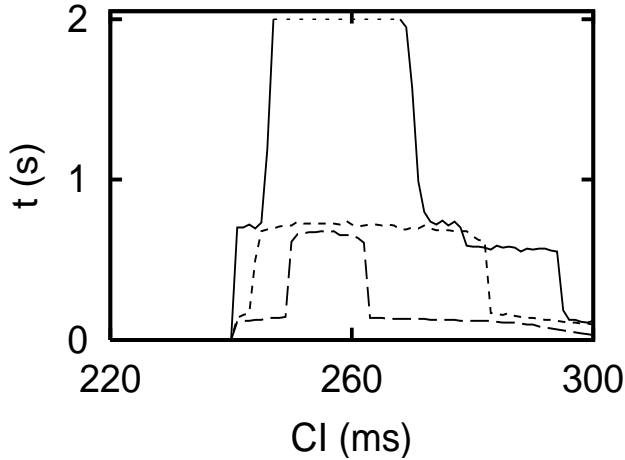}
\caption{
\label{CIlifetime}
The length of reentry as a function of coupling interval for the transmural sheet with 
$\mu=0.08$, basic cycle length=1000 ms, $w=0.32$ cm  and 
three different values of $r$ (0.32cm (solid), 
0.24cm (short dashed) and 0.16cm (long dashed)). 
Reentrant excitations for $r=0.32$ cm (solid line) that were
characterized by a pinning to the bubble and that persisted throughout the duration of 
our numerical simulations (2 s) are indicated by the dotted line.
}
\end{center}
\end{figure}

Our main result is summarized in Fig. \ref{2dphase} where we
show the length of reentry, measured in seconds, 
as a function of the two geometry parameters, $w$ and $r$,
for fixed long QT syndrome strength $\mu=0.08$. 
Of course, the duration of reentry is also dependent on the 
coupling interval (see below and Fig. \ref{CIlifetime}) and to
determine its value, we varied the coupling interval
in steps of 1 ms and measured the 
duration of reentry for the five coupling intervals 
that produced the longest lived reentrant excitations.
The average of these five events   
is shown in Fig. \ref{2dphase} using a gray scale with white 
corresponding to short-lived reentry and black corresponding to long-lived
reentry. 
Note that the diagonal corresponds to a strip with a rounded edge.

\begin{figure}
\begin{center}
\includegraphics{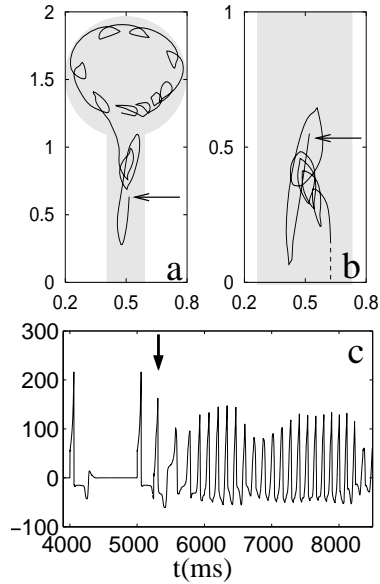}
\caption{
\label{tip}
Spiral tip trajectories for
$w=0.48$ cm and $r=0.24$ cm (a) and $w=0.24$ cm, $r=0.32$ cm (b).
The M cell domain is lightly shaded and the long QT syndrome strength is
$\mu=0.08$.
In both cases, the start of the tip trajectory is indicated
by an arrow and the initial stage of the reentrant
excitation is not shown. Note that only a part of the 
computational domain is shown.
In (a), the dashed line is used as a guide to the eye in
order to show that the spiral ends up leaving the sheet.
The size of the tissue sheet is 1 cm $\times$ 2.5 cm.
(c) Simulated pseudo-ECG corresponding to the reentry shown
in (b), showing the 4th and 5th regular beat and the 
premature stimulus indicated by the arrow.}
\end{center}
\end{figure}

From the figure, one can see that reentry that
lasts longer than 1 s requires a sufficiently large M cell domain.
This can be accomplished by either having a 
wide strip ($w>0.48$ cm), or having 
a large bubble ($r>0.28$ cm for $w=0.24$ cm).
These two cases, however, result in 
qualitatively different reentrant patterns.
A typical spiral tip trajectory 
for a wide, rounded-edged, strip is shown in Fig. 
\ref{tip}a.
The tip was calculated every 1 ms using the intersection of the iso-contour
$V=-30$ mV and the line defined by $h=0.5$, where $h$ is the 
inactivation gate of the sodium channel\cite{Claetal01}.
The initial phase of the reentry is not shown and the first 
point of the tip trajectory is marked by the arrow.
The spiral eventually drifts out of the computational domain, which 
is indicated by the dashed line. 
The total trajectory in Fig. \ref{tip}a represents  
1 s and approximately 5 spiral rotations.

\begin{figure}
\begin{center}
\includegraphics{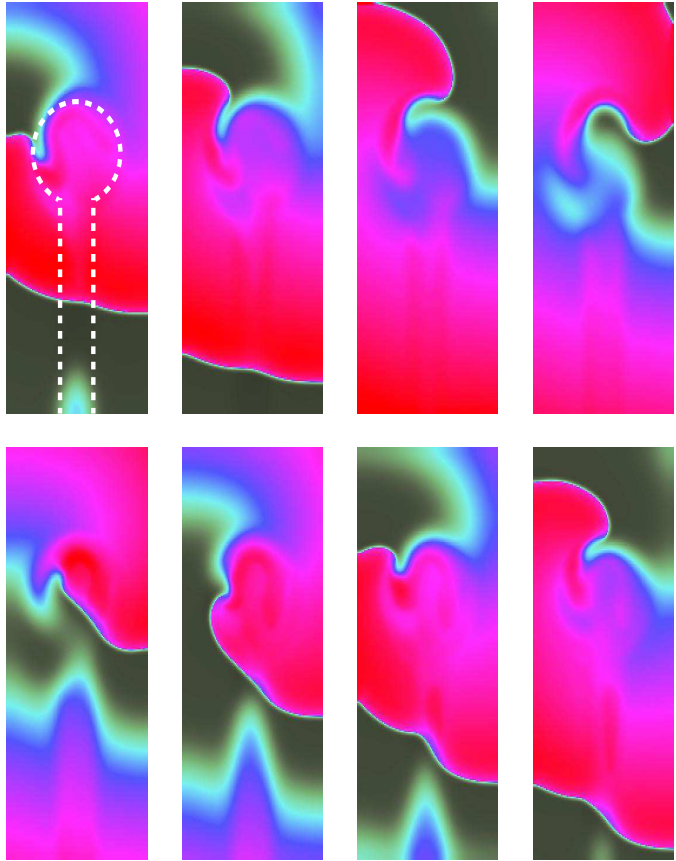}
\caption{
\label{2dcolor} 
Color plots of the membrane potential  of the reentrant 
excitation of Fig. \ref{tip}b. 
The M cell domain is indicated by the white dashed line.
The plots are taken 25 ms apart and start 
1s after the application of the premature stimulus.
The first one is  
in the upper left hand corner 
while the last one is in the lower right hand corner. 
The color scale is identical to the one employed in Fig. \ref{spacetime1d}.
Parameter values: $\mu$=0.08, $w=0.24$ cm, $r=0.32$ cm and 
a physical domain of  1 cm $\times$ 2.5 cm}
\end{center}
\end{figure}

The tip trajectory in the case of a large bubble is shown in 
Fig. \ref{tip}b where we again have omitted the initial part of the reentrant 
excitation and marked the first point of the trajectory with an arrow. 
Here, the spiral tip becomes pinned to the bubble,
leading to long lasting-reentry. 
A qualitative similar picture was found in a numerical 
study of a tissue sheet with a central ischemic region \cite{Claetal02}.
The spiral tip is plotted for roughly 2 s and 10 spiral rotations 
in Fig. \ref{tip}b.
The corresponding simulated pseudo-ECG, using the algorithm in 
\cite{PloBar88}, is shown in Fig. \ref{tip}c.
In this figure, we have included the two 
regular preceding beats  and have indicated the premature 
stimulus by an arrow.
The polymorphic undulating ECG is suggestive of torsade de pointes.
Finally, the reentrant excitation of Fig. \ref{tip}b is also plotted in 
Fig. \ref{2dcolor} where we show
the activation of the tissue 
every 25 ms using the color scale of Fig. \ref{spacetime1d}.

Since reentry is caused by the partial blocking of the premature
stimulus, we found a good correlation between the range of coupling 
intervals that leads to reentry and the range of coupling intervals that 
leads to a 1D conduction block. 
This is illustrated in Fig. \ref{CIlifetime} where we have 
plotted the  length of reentry as a function of the 
coupling interval for $\mu=0.08$, $w=0.32$ cm
and three different values of $r$. 
For these parameter values, 
the 1D cable simulations reveal 
a conduction block for coupling intervals
between 240 ms and 300 ms. 
In 2D, the onset of reentry occurs for a coupling interval of 240 ms, as can 
be seen in Fig. \ref{CIlifetime} and no reentry is observed for coupling intervals larger than 300ms. As already mentioned before, 
a larger bubble size  leads  to a more sustained reentry.
In fact, for large bubbles, the spiral tip  becomes attached to 
the bubble, resulting in reentry persisting throughout the full duration  of
 our numerical simulations (2 s). 
These points are indicated by the dotted line in Fig. \ref{CIlifetime}. 

\section{Discussion}
We have presented a numerical study of the role of a particular type of
spatial heterogeneity in cardiac tissue in the initiation and maintenance
of reentrant excitations. 
Specifically, we have addressed the role of M cells, located 
in the midmyocardium. The prolonged action potential of M cells leads to 
dispersion of repolarization which is increased significantly in patients 
with the long QT syndrome.
We have focused on the long QT 3 syndrome 
which is marked by an increase in the  late \ina\ and have 
investigated two previously proposed mechanisms, using 
numerical simulations of inhomogeneous tissue cables and  sheets.

The first mechanism proposes that early after depolarizations in the M cell domain leads to 
premature stimuli in surrounding tissue. 
However, our numerical results show that under normal electrotonic 
coupling, the late sodium current is not sufficiently strong to 
induce an excitation in neighboring points.
Typical currents that are responsible for the 
generation of early after depolarizations are fairly 
small \cite{LuoRud94b}.
For example, in the case studied here,  the late \ina\ current is 
on the order of 2-4 $\mu$Acm$^{-2}$ while in long QT 1 and long QT 2,  
the respective reduction of \iks\ and \ikr\ is 
on the order of 1 $\mu$Acm$^{-2}$.
Furthermore, the L type calcium current involved in the generation
of early after depolarizations is roughly of the same magnitude.
This 
should be compared to the strength of the injection current needed to induce
a counter propagating wave in a cable with $\mu=0$. 
Numerically, we find that the necessary injection current 
is approximately an order of magnitude larger 
($\sim30 \, \mu \rm{A} \, cm^{-2}$).
It is therefore
perhaps not a surprise that the inclusion  of a leak
current in \ina\ 
is not sufficient to create a wave propagating in the
unidirectional direction. 
We have verified that a reduction of \ikr\
was also unable to initiate unidirectional propagation.

We also investigated the effect of a discontinuity in conductivity
between the M cell domain and the epicardium, 
as reported in a study \cite{Yanetal98}.
We found that a large increase in resistance between the two domains
could result in an extra stimulus in the epicardium that was
provoked by an early after depolarization in the M cell domain. 
However, for the model used in our study, this increase 
was at least four times higher that the one reported in the 
literature. 
This indicates that the discontinuity in conductivity 
between the M cell domain and epicardium 
is likely insufficient to generate a premature stimulus arising 
from the M cell domain.
 
Our simulation study focused on early after depolarizations
originating in M cells. 
We did not consider other possible sources of 
early after depolarizations. 
For example, cells in the right ventricular epicardium exhibit
a prominent transient outward current $I_{\rm{to}}$ \cite{DiDetal96}.
Early after depolarizations originating from the 
right ventricle epicardial cells might 
play an role in phase 2 reentry \cite{LukAnt96}, which is thought to
be associated with the Brugada syndrome \cite{Antetal03}. 
A recent numerical study,
using a modified version of the  LRd model that included  
$I_{\rm{to}}$ together with a large reduction
in the time constant for the L type calcium channel,
investigated this scenario \cite{Miyetal03}.
These modifications were found, under very specific conditions, 
to elicit a secondary wave in a cable divided 
into two subdomains with different $I_{\rm{to}}$ densities.
However, this scenario is specific to 
right ventricular epicardial cells and is unlikely to play a role in 
reentry phenomena associated with M cell domains in the left 
ventricle.

The second mechanism we investigated
relies on the slow repolarization of the M cell domain
which functions as a wave block for subsequent stimuli. 
When the M cell domain is inhomogeneous, or alternatively, 
if the pacing breaks the translational symmetry along the endocardial 
wall, a properly timed extra stimulus can be blocked in one area
of the myocardium while allowed to propagate in another. 
This, then, could create reentrant excitation. 

We first studied the propagation of stimuli, delivered prematurely
with varying coupling intervals, for different basic cycle length and 
long QT syndrome strength. We found a rich set of possible
dynamics of the wave following the premature stimulus. 
Particularly noteworthy  was the 
dramatic reduction of the wave speed 
occurring within the M cell domain.
Prerequisite for this behavior is a sufficiently long  basic cycle length.
Preliminary investigations of the dynamics indicate that during this
slow propagation the usual sodium channel activation of 
neighboring cells fails. 
Instead, the calcium channel becomes responsible for
the depolarization and the spread of the impulse.
This scenario has also been observed experimentally in 
growth cultures of rat myocytes \cite{RohKuc97}.

The 1D simulations revealed two separate regions in the 
basic cycle length vs. $\mu$ space in which conduction block could occur.
The first region, characterized by relatively small basic cycle length,
exhibits a small conduction block window.
For coupling intervals within this window, the wave fails to 
propagate abruptly.
The second region, characterized by a slow wave propagation 
and a subsequent wave block, has a much larger conduction block window.
Furthermore, along the boundary of this region,
the dynamics is  
not straightforward. A window of coupling intervals that exhibit 
conduction block is followed by a window of coupling intervals that show 
slow propagation within the M cell domain.
Again, the calcium handling seems to be a likely candidate for 
an explanation of this behavior.
This is supported by recent experimental     
work, able to visualize voltage and calcium simultaneously,
that showed that
calcium dynamics plays a crucial role in the 
generation of early after depolarizations in
particular and action potential prolongation in general 
\cite{Choetal02}.
Thus, a model that faithfully incorporates calcium handling
is essential.
This might require a description of local spatial events 
including calcium sparks \cite{Bers_book01}, 
currently absent from the LRd model.

In mapping out the phase diagram of Fig. \ref{2dphase} we have 
focused on a point in the 
parameter space within this second region  (basic cycle length of 
1000 ms and $\mu$=0.08).
The  main finding of our 2D simulations is that the second
mechanism can lead to reentrant excitations. 
Indeed, for large enough $w$, or large enough $r$, we 
find reentry that can last 1 s or more. 
The observed pseudo-ECG, with its polymorphic undulating morphology, 
is suggestive of torsade de pointes. 
Furthermore, for a large bubble size, the reentry becomes pinned to 
the region of heterogeneity, thus leading to long lasting
reentrant excitations.

We should note that the observed mechanism for reentrant excitations 
is not specific to long QT 3.
We have also investigated both the first and second mechanism in 
a model for long QT 2. 
In this model, we reduced the maximal conductance of  \ikr\
which  allowed us to vary the strength of the long QT syndrome. 
As in our long QT 3 model, we found that early after depolarizations did occur in the M cell domain
but were not able to excite neighboring tissue. 
On the other hand, when an M cell domain was included in the 
transmural sheet, the long QT 2 model was able to partially block 
a premature stimulus and sustained reentrant excitations were found. 
Also, we point out that we have replicated the pacing protocol
employed in Ref. \citen{Akaetal02}. 
There, the endocardium was paced at a long basic cycle length, followed by a 
premature stimulus on the epicardium. 
As the pacing protocol already breaks the translational symmetry 
along the endocardium, we included a continuous strip in our 
numerical geometry. 
We found that this pacing protocol could lead to reentry, 
provided that the strip had a sufficient width.  
Besides the geometrical construction of this paper or the pacing protocol
of \cite{Akaetal02}, 
there are other ways to break the 
translational symmetry.
For example, a strip of M cells with domains of different long QT syndrome strength 
can have, for a given coupling interval, domains that exhibit wave block while
other domains support normal propagation.
This, again, would result in partial wave block and subsequent reentry.

The mechanism that was shown here to create reentrant excitation
relies on an extra stimulus, delivered at a relatively small 
coupling interval. 
This extra stimulus can possibly be generated by an
early after depolarization in a Purkinje fiber that successfully 
excites the endocardium. 
A possible mechanism that would lead to 
the initiation of a wave originating from a Purkinje fiber
early after depolarization is based on the 
experimentally observed  reduced coupling between 
the Purkinje fiber and the endocardium\cite{RawJoy87}.
Under  this condition, 
numerical simulations have revealed that the inclusion of 
a large resistive barrier between the Purkinje fibers and 
the ventricular cells can lead to
early after depolarizations in the Purkinje fibers that
are able to 
generate extra stimuli in the ventricular domain\cite{Monetal00}.

Our numerical simulations were carried out using 
a simplified transmural geometry 
of the ventricular wall that consisted of  a centrally located
M cell domain, sandwiched between two electrophysiologically 
equal domains representing the endo- and epicardium.  
Incorporating more realistic geometries and endo- and 
epicardial cells with different electrophysiological properties,
however, will most likely not change the qualitative conclusions
of this study.  
Furthermore, the action potentials of the 
epicardial and endocardial cells recorded in the 
wedge preparation of the canine left ventricle, 
were very similar \cite{Akaetal02}.

The numerical simulations employed 
2D transmural sheets while the actual heart is of course a 
much more complicated 3D object. 
Nevertheless, our numerical work, 
replicating the  experiments in wedges \cite{Akaetal02},
demonstrates that the creation of an intramural reentry event 
is plausible under certain conditions. 
In 3D, an  ensuing intramural filament can either close 
on itself, leading to a scroll ring that remains intramural and is on the  
scale of the ventricles 
\cite{Elsetal96, Veretal97}, or can reorient and attach itself
to the endo- and epicardium. 
Possible subsequent instabilities could destabilize this filament,
resulting in multiple filaments and fibrillation.
Only full-scale 3D simulations, preferably on realistic
geometries containing detailed anatomical information, 
can address these events.
We are currently investigating this possibility, which 
includes  the development of novel algorithms that can 
accurately simulate electrical wave propagation in 
anatomically realistic hearts \cite{Cheetal03}.
 
\vspace{1cm}
\noindent
The authors acknowledge support from the Whitaker 
Foundation and 
from the NSF sponsored Center for Theoretical Biological Physics 
(grants PHY-0216576 and 0225630).

\end{document}